\documentclass[reprint,aps,prb]{revtex4-2}
\usepackage{chemformula} % Formula subscripts using \ch{}
\usepackage[T1]{fontenc} % Use modern font encodings
\usepackage{amsmath,amssymb,graphicx,subfigure,braket,siunitx,url,booktabs,multirow,appendix}
\usepackage[caption=false]{subfig}
\usepackage{caption}
\usepackage{dcolumn}% Align table columns on decimal point
\usepackage{bm}
\newcommand{\ii}{\mathrm{i}}

\begin{document}

\preprint{}

\title{Electron correlations in kagome metals \ch{AV3Sb5}  (A= K, Rb, Cs)}

\author{Feihu Liu}
\email[{Corresponding author: }]{liufeihu@xupt.edu.cn}

\altaffiliation{}
\affiliation{School of Sciences, Xi’an University of Posts and Telecommunications, Xi’an 710121, P. R. China}

\author{Changxu Liu}
%\email[]{liuchangxu04@163.com}
\affiliation{School of Sciences, Xi’an University of Posts and Telecommunications, Xi’an 710121, P. R. China}

\author{Maolin Zeng}
%\email{zengmaolin@stu.xupt.edu.cn}
\altaffiliation{}
\affiliation{School of Sciences, Xi’an University of Posts and Telecommunications, Xi’an 710121, P. R. China}

\author{Qiyi Zhao }
%\email[]{qiyi_xiyouphy@163.com}
\affiliation{School of Sciences, Xi’an University of Posts and Telecommunications, Xi’an 710121, P. R. China}

\date{\today}

\begin{abstract}
The investigation of electronic order-quantum phase interplay in Kagome lattices commonly employs the extended Kagome-Hubbard model, where the critical parameters comprise on-site $(U)$ and intersite $(V)$ Coulomb interactions. In prototypical kagome metals \ch{AV3Sb5} (A = K, Rb, Cs), the geometrically frustrated quasi-2D architecture induces pressure-dependent complexity in vanadium d-electron correlations, necessitating systematic theoretical scrutiny.
Utilizing the $d-dp$ model within constrained random phase approximation (cRPA), we quantified $U$, $V$, and Hund's coupling $J$ under hydrostatic pressure (0-9 GPa). While \ch{KV3Sb5} and \ch{RbV3Sb5} exhibit pressure-insensitive interaction parameters, \ch{CsV3Sb5} manifests anomalous discontinuities in $U$ and $V$ near $0.2$ GPa, suggesting a first-order electronic phase transition. This work establishes cRPA-derived interaction landscapes as critical predictors for pressure-tunable quantum phenomena in correlated kagome systems, offers a new insight into the understanding of the interplay between the CDW transition and the double superconductivity dome in \ch{CsV3Sb5} at low pressure.
\end{abstract}

%\keywords{Superconductivity}

\maketitle
\section{Introduction}
\label{sec1}

The vanadium-based kagome metals \ch{AV3Sb5} (A= K, Rb, Cs) \cite{ortiz_new_2019,ortiz_cs_2020,neupert_charge_2022,wilson_av3sb5_2024} have emerged as a prototypical platform exhibiting extraordinary diversity in correlated electronic phenomena. Particularly noteworthy is the observation of time-reversal symmetry breaking (TRSB) in the charge density wave (CDW) state \cite{xu_three-state_2022,wu_simultaneous_2022,hu_time-reversal_2022,wang_electronic_2021,shumiya_intrinsic_2021,jiang_unconventional_2021,yu_concurrence_2021,mielke_time-reversal_2022}, which intriguingly coexists with potential unconventional superconductivity \cite{wu_nature_2021,kiesel_sublattice_2012,zhou_kagome_2024}. Systematic pressure-dependent studies have elucidated the complex interplay between these two competing orders in \ch{AV3Sb5} compounds, with mounting evidence pointing to their interwinded relationship \cite{guguchia_tunable_2023,zhou_kagome_2024,yu_unusual_2021,chen_double_2021,du_pressure-induced_2021,wang_competition_2021,du_evolution_2022,li_discovery_2022,zhu_double-dome_2022}. This pressure-tuned competition manifests most dramatically in the distinct phase diagrams of these isostructural compounds, suggesting subtle yet crucial differences in their electronic correlation landscapes.

Systematic pressure-dependent measurements of the charge density wave transition temperature $(T_{CDW})$ and superconducting critical temperature $(T_c)$ reveal their competitive interplay through distinctive dome-shaped profiles in the $p-T$ phase diagram. In \ch{KV3Sb5}, $T_c$ undergoes dramatic enhancement from $\approx 0.9$ K at ambient pressure to $3.1$ K at $p\approx 0.4$ GPa, while $T_{CDW}$ is progressively suppressed, vanishing completely at a critical pressure $p\approx 0.5$ GPa. Subsequent pressure increases up to $9$ GPa lead to a monotonic suppression of superconductivity, forming a singular superconducting dome \cite{du_pressure-induced_2021}. Notably, \ch{RbV3Sb5} exhibits a contrasting M-shaped double-dome structure in its $T_c(p)$ evolution, with two distinct extreme $T_c \approx 4.4$ at $p_{c1} \approx 1.5$ and $T_c \approx 3.9$ K around $p_{c1} \approx 1.5$ \cite{wang_competition_2021}. 

The superconducting phase diagram of \ch{CsV3Sb5} exhibits a markedly more pronounced double-dome structure compared to \ch{RbV3Sb5}, with two distinct superconducting transitions emerging at critical pressures $p_{c1} \approx 0.7$ GPa and $p_{c2} \approx 2$ GPa. This pressure-dependent evolution suggests a more intricate interplay between CDW ordering and superconductivity in \ch{CsV3Sb5} \cite{yu_unusual_2021}. Transport property analyses, including magnetoresistance measurements and residual-resistivity ratio quantification, associate both pressure-induced transitions with successive modifications of the CDW state \cite{yu_unusual_2021}. These findings imply the potential coexistence or competition between two distinct charge-ordered phases and superconducting order in \ch{CsV3Sb5} under applied hydrostatic pressure.

Current theoretical frameworks investigating competing charge orders in \ch{AV3Sb5} compounds have predominantly focused on the extended Kagome-Hubbard model (KHM), where both on-site Hubbard U and nearest-neighbor Coulomb interaction V prove crucial for determining the system's phase behavior \cite{kiesel_sublattice_2012,denner_analysis_2021,profe_kagome_2024,ferhat_phase_2014,ferrari_charge_2022,kiesel_unconventional_2013}. Systematic investigations employing Ginzburg-Landau formalism have mapped the U-V parameter space, revealing a phase boundary between charge bond order (CBO) and charge density order (CDO) that depends critically on the $U/V$ ratio \cite{denner_analysis_2021}. 

Recent advancements in functional renormalization group studies of the generalized KHM framework \cite{profe_kagome_2024} (decoupled from specific material parameters) further unveil a rich phase diagram containing multiple charge-ordered states. Additionally, a novel CDW order, the chiral flux order, has recently been proposed \cite{2021SciBu..66.1384F,PhysRevB.104.165136}. Featuring a circulating current pattern, this phase breaks time-reversal symmetry and is identified as a topological state. It is considered a key factor enabling topological superconductivity and other phenomena \cite{2023PhRvB.107f4506J,PhysRevB.109.104512,PhysRevB.110.134502}. These theoretical developments highlight the critical role of electronic correlations at different van Hove singularities in governing competing charge orders within the Kagome lattice system. 

Collectively, these considerations underscore the paramount importance of precise quantification of Coulomb interaction parameters ($U$, $V$) in understanding the complex electronic landscape of \ch{AV3Sb5} systems. Our study addresses this critical need through constrained random phase approximation (cRPA) calculations that systematically track the pressure-dependent evolution of these interaction parameters. This systematic investigation provides crucial insights into two unresolved questions in the field: (1) the absence of a low-pressure double superconducting dome in \ch{KV3Sb5} compared to its cesium counterpart, and (2) the fundamental mechanisms driving CDW transitions in \ch{CsV3Sb5}. The derived pressure-$U$-$V$ correlations establish a quantitative framework for reconciling observed material-specific differences in electronic ordering phenomena across the \ch{AV3Sb5} family.

\section{Computation methods}
In solids, at the lowest approximation, namely the random phase approximation (RPA), the  screened Coulomb interaction is given by
\begin{equation}
  \label{eq:full}
  W=(1-vP)^{-1}v,
\end{equation}
where $v$ is the bare Coulomb interaction and $P$ is the polarization operator given by 
\begin{equation}
  \label{eq:polarization}
  \begin{split}
    P(r,r^\prime;\omega)  &=\sum_i^{occ} \sum_j^{unocc} \psi_i(r)\psi_i^\star(r^\prime)\psi_j^\star(r)\psi_j(r^\prime)\\
    &\times \left\{\frac{1}{\omega-\epsilon_j+\epsilon_i+\ii 0^+} -\frac{1}{\omega+\epsilon_j-\epsilon_i-\ii 0^+}\right\},
  \end{split}
\end{equation}
where $\psi_i$ and $\epsilon_i$ are the Kohn-Sham eigenfunctions and eigenvalues obtained from DFT calculation. Summations over indices $i$ and $j$ run over occupied and unoccupied states, respectively.

If we consider the V-$3d$ bands as a correlated subsystem, the polarization may be divided into two parts $P=P_r+P_d$, in which $P_d$ only include screening from the V-$3d$ orbitals, and $P_r$ is the rest of the polarization. It is straightforward to calculate $P_d$ by restricting the summation in Eq.\eqref{eq:polarization} to V-$3d$ bands only, namely the cRPA method. Then the effective interactions among electrons living in V-$3d$ bands are given by
\begin{equation}
  \label{eq:u}
  U(\omega) = (1-vP_d)^{-1}v.
\end{equation}
Note that, the definition of $U(\omega)$ in Eq.\eqref{eq:u} is independent of the basis function and is a general approach. However, in realistic system, it is often complicated by the issue of correctly constructing the correlated subsystem. 

\begin{figure*}[!ht]
\centering
\includegraphics[width=0.8\textwidth]{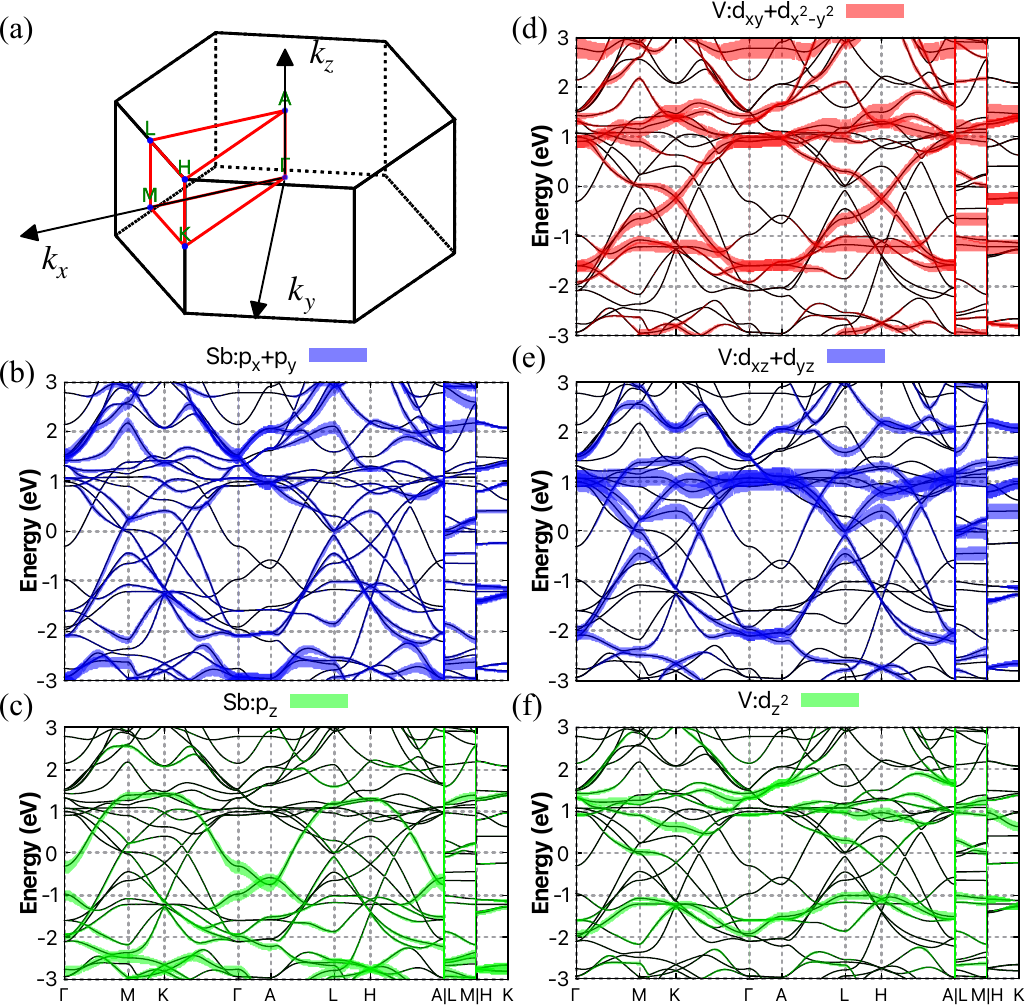}
\caption{The orbital projected bands of \ch{CsV3Sb5}. The V-$d$ dominant bands [(d)-(f)] exhibit characteristic flat-bands behavior of two-dimensional kagome lattice. Conversely, the Sb-$p$ dominant bands [(b)-(c)] display strong three-dimensional character, evidenced by significant band dispersion along $k_z$ direction. Comparing (b) and (e), the overlap of Sb-($p_x+p_y$) and V-($d_{xz}+d_{yz}$) orbitals is also noticeable.}
\label{fig:pjbands}
\end{figure*}

As a typical example, in \ch{CsV3Sb5}, the V-$3d$ bands that spanned the correlated space are not well isolated from the Sb-$5p$ bands, which can be seen from the projected band structure depict in FIG.~\ref{fig:pjbands}. How to uniquely identify the $d$-$d$ screening channel to obtain the polarization $P_d$ is a major issue in cRPA method. This involves searching for some well-defined localized states to be a basis set. One popular choice is the maximally localized wannier functions (MLWF) $\ket{\phi_{Rm}}$, which can be calculated by the Fourier transformation from the reciprocal to the direct space of DFT Kohn-Sham eigenfunctions \cite{marzari_maximally_2012}. 

In this regard, there are four different approaches to calculate the polarization: (a). The band method \cite{aryasetiawan_frequency-dependent_2004} which directly uses Adler-Wiser equation \cite{adler_quantum_1962,wiser_dielectric_1963} to calculate $P_d$. This method only works if the target space is well isolated from the rest and there is no entanglement between them. (b). The disentanglement method which heavily relies on the special choice of a energy window within which the Wannier basis can be constructed so that the rest and the correlated subspace are fully decoupled. However, in some cases, the strongly dependence of the energy window will cause unnecessary arbitrariness \cite{miyake_ab_2009}. (c). The weighted method introduces probability weights of KS functions being in the target Wannier space to define the polarization \cite{sasioglu_effective_2011}. (d). The projector method, which is similar to the weighted method, uses projector between Bloch functions and Wannier functions to construct target space spanned by correlated Bloch functions. This method uses k-p perturbation theory to calculate the long-wave limit contribution which is ill defined for metallic system \cite{kaltak_merging_2015}. As in our case, the long-wave limit will not be included for the projector method.

\begin{figure}[t]
\centering
\includegraphics[width=0.4\textwidth]{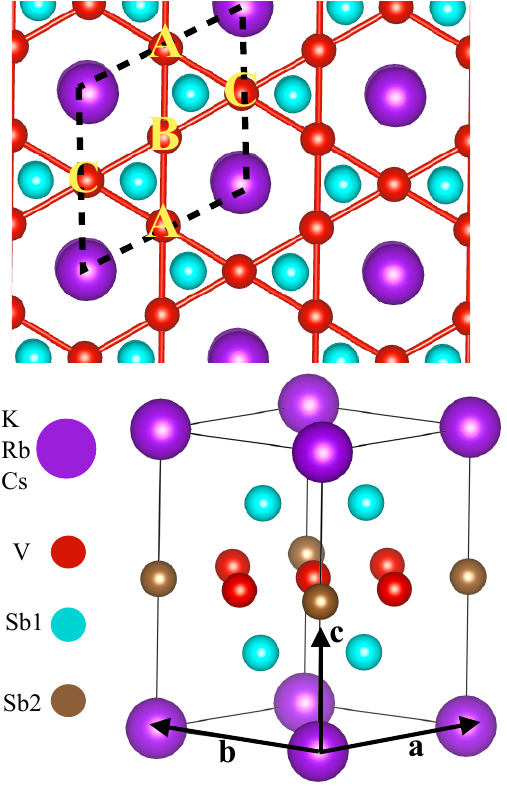}
\caption{Crystal structure of \ch{AV3Sb5}. The dotted line indicates a unit cell, in which three atomic V sites are labeled by $A,B,C$.}
\label{fig:structure}
\end{figure}

We further calculate the effective interaction matrix elements defined by 
\begin{equation}
  U_{m_1m_2m_3m_4}^{R_1R_2R_3R_4}(\omega)=\bra{\phi_{R_1m_1}\phi_{R_2m_2}}U(\omega)\ket{\phi_{R_3m_3}\phi_{R_4m_4}},
\end{equation}
where $R_i$ labels the atomic sites and $m_i$ refers to angular quantum numbers indicating the dominant orbital character of wannnier funciton. For \ch{AV3Sb5}, there are three \ch{V} atoms in one unit cell, one can easily obtain the on-site, nearest site interaction $U$ and $V$ respectively:
\begin{equation}
  \begin{split}
      U_{m_1m_2m_3m_4}(\omega) &= U_{m_1m_2m_3m_4}^{R_1R_2R_3R_4}(\omega)|_{R_1=R_2=R_3=R_4}\\
      V_{m_1m_2m_3m_4}(\omega) &= U_{m_1m_2m_3m_4}^{R_1R_2R_3R_4}(\omega)|_{R_1=R_2=A;\,R_3=R_4=B}
  \end{split}
\end{equation}
where {A, B, C} label three V sites schematically displayed in Figure \ref{fig:structure}.

For practical usage within many-body techniques via tight-binding Hamiltonian, it is convenient to further reduce these matrices elements to a reasonable number of parameters, such as the orbital averaged onsite Coulomb interaction $U$, nearest site interaction $V$ and the Hund exchange $J$:
\begin{equation}
  \label{eq:hk}
  \begin{split}
      U &=\frac{1}{N}\sum_{m=1}^N U_{mmmm}\\ 
      V &=\frac{1}{N}\sum_{m=1}^N V_{mmmm}\\
      J &=\frac{1}{N(N-1)} \sum_{m\ne m^\prime}^N U_{mm^\prime m^\prime m},
  \end{split}
\end{equation}
where $N$ the number of orbitals. Note that the calculated interaction parameters in Eq.\eqref{eq:hk} are genuinely depending on the frequency. However,  in our comparative analysis to the study based on extended Kagome-Hubbard mode, in which the frequency dependence of the Hubbard U parameter was not accounted for. Consequently, only the static $U$ at $\omega=0$ is considered in our calculation.

\begin{figure*}[ht!]
\centering
\includegraphics[width=0.8\textwidth]{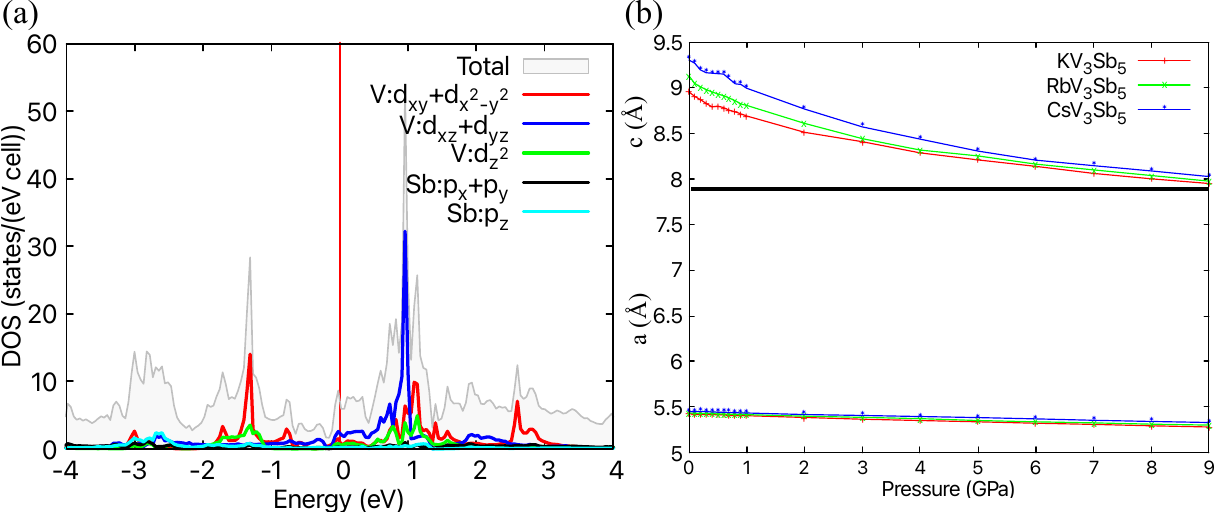}
\caption{(a) Projected density of states of \ch{CsV3Sb5}. (b) The lattice parameters of \ch{AV3Sb5} as function of pressure. The V-Sb layer distance decreases more aggressively upon increasing pressure due to the quasi-2D nature of \ch{AV3Sb5}. }
\label{fig:pdos}
\end{figure*}

In cRPA calculations of Coulomb interaction parameters, a key step is selecting the correlated subspace and obtaining its polarizability. Here, crystal field effects separate the V-3d bands into three distinct groups: $\{d_{xy}+d_{x^2-y^2}\}$, $\{d_{xz}+d_{yz}\}$ and $\{d_{z^2}\}$. While any of these could, in principle, be chosen as the correlated subspace, significant orbital overlap is evident in the projected bands (FIG.~\ref{fig:pjbands}) and projected density of states (PDOS, FIG.~\ref{fig:pdos}(a)). Consequently, we define the total V-$3d$ space as the correlated subspace.

The next step involves constructing localized atomic Wannier functions from a set of bands. Although various methods exist, an optimal choice must satisfy two criteria: the Wannier function spread should be reasonably small, and key features of the electronic structure, such as van Hove singularities near the Fermi level and the Fermi surface topology, must remain unperturbed. Orbital-projected bands (FIG.~\ref{fig:pjbands}) reveal that Sb-$5p$ orbitals contribute to the van Hove singularities at the $M$ point (see also Ref.~\cite{jeong_crucial_2022} for details on the role of Sb-$5p$ orbitals).

\begin{figure*}[ht!]
\centering
\includegraphics[width=0.8\textwidth]{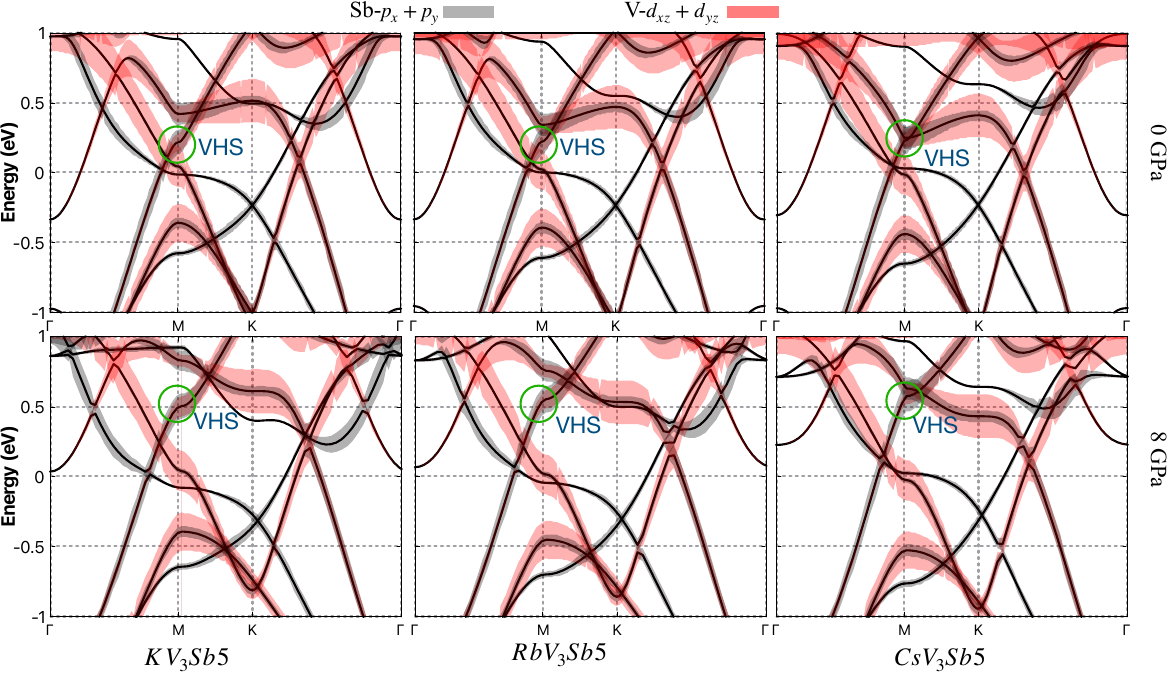}
\caption{ The VHS (marked by a green circle) formed by Sb-($p_x+p_y$) and V-($d_{xz}+d_{yz}$) orbitals respond to pressure. With increasing pressure, the energy position of the VHS is moving toward higher energy.}
\label{fig:vhs}
\end{figure*}

Furthermore, FIG.~\ref{fig:pjbands} also exhibits substantial overlap between Sb-($p_x+p_y$) and V-($d_{xz}+d_{yz}$) bands. More detailed comparisons across the \ch{AV3Sb5} compounds (A = K, Rb, Cs) reveal pressure-dependent behavior. FIG.~\ref{fig:vhs} shows the response of hybrid Sb-($p_x+p_y$)/V-($d_{xz}+d_{yz}$) bands to pressure, with the green cirlce indicating an important Von Hove singularity (VHS) for which Sb-$5p$ takes a large portion. This VHS is presumably related to the Fermi surface nesting according to the experiments \cite{kang_twofold_2022} and its energy position is correlated to CDW instability \cite{PhysRevB.104.144506}. Importantly, the energy level of this VHS is sensible to the pressrue because of its Sb-$5p$
character \cite{PhysRevB.104.205129,jeong_crucial_2022}. This observation suggest that the inclusion of Sb-$5p$ bands is essential to study Coulomb interaction in Kagome metal. Consequently, the Wannierization energy window encompasses all Sb-$5p$ and V-$3d$ bands, while the screening channel remains restricted to V-$3d$ orbitals. This corresponds to the $d$-$dp$ model described in Refs.~\cite{vaugier_hubbard_2012,panda_pressure_2017}.

\begin{figure*}[ht!]
\centering
\includegraphics[width=0.9\textwidth]{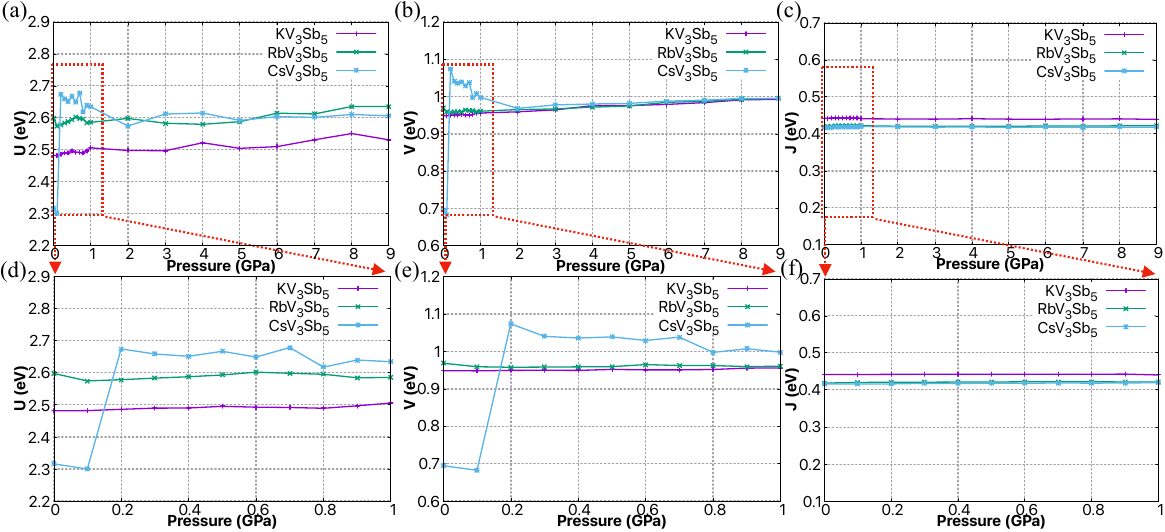}
\caption{The evolution of (a) $U$, (b) $V$ and (c) $J$ as function of pressure respectively. (d)-(f) are the enlarged views of the low-pressure section.}
\label{fig:uv}
\end{figure*}

\emph{Numerical details.} All DFT calculations are performed by using VASP package \cite{kresse_ab_1993,kresse_efficiency_1996,kresse_efficient_1996}, in which the generalized gradient approximation \cite{perdew_generalized_1996} with the PBEsol exchange-correlation functional \cite{perdew_restoring_2008} are implemented.  Due to the metallic nature of our system, the Methfessel-Paxton smearing technique was employed \cite{PhysRevB.40.3616}. Starting from the experimental lattice parameters and atomic positions \cite{ortiz_new_2019}, the crystal structure and internal coordinates were relaxed. To ensure more accurate calculations of total energy and forces, this relaxation used a denser k-mesh density of $11\times 11 \times 7$, a plane-wave basis energy cutoff of 600 eV, and included van der Waals dispersion corrections via the DFT-D3 method \cite{grimme_consistent_2010}. External hydrostatic pressure was applied by setting the stress tensor. The localized Wannier functions are constructed by VASP2WANNIER90 interface and WANNIER90 package \cite{pizzi_wannier90_2020,mostofi_updated_2014}. In cRPA and the GW bands calculation, increasing the k-point density dramatically increases the computational cost. According to the convergence check in the Appendix (see FIG.~\ref{fig:convergency}), we choose a optimal $9\times 9 \times 6$ k-mesh and $107$ empty bands.

\section{Results}
The evolution of screened $U$, $V$ and $J$ as a function of pressure $p$ is plotted in FIG.~\ref{fig:uv}. In \ch{KV3Sb} and \ch{RbV3Sb}, all $U$, $V$ and $J$ are nearly fixed except for \ch{CsV3Sb}, in which $U$ and $V$ exhibit a dramatic fluctuation in the low pressure region ($p<1$ GPa). It's also worth noting that \ch{KV3Sb} has the lowest $U$ and highest $J$ in the high pressure region ($p>1$ GPa), while \ch{RbV3Sb} and \ch{CsV3Sb} have almost identical $U$, $V$ and $J$. On the contrary, the bare Coulomb interactions exhibits no sudden changes (see FIG.~\ref{fig:bare} in the Appendix), meaning that the anomalous changes in the effective Coulomb interaction ($U$ ,$V$) arise from the screening effect.

To comprehensively investigate the anomalous electronic fluctuations in \ch{CsV3Sb5}, we extended our analysis beyond the low-pressure regime. As illustrated in FIG.~\ref{fig:pdos} (a), the structural response of \ch{AV3Sb5} under pressure reveals a critical dichotomy: while the in-plane lattice parameter $a$ remains nearly invariant, the interlayer spacing along the $c$-axis undergoes monotonic compression. This decoupling motivates our strategy of isolating the $c$-axis modulation as a virtual pressure proxy, enabling systematic exploration of extended interlayer distances beyond ambient conditions $(c > c_0)$. Through this controlled dimensional tuning, we mapped the complete evolution of Coulomb parameters across the critical low-pressure landscape.
\begin{figure*}[ht!]
\centering
\includegraphics[width=0.7\textwidth]{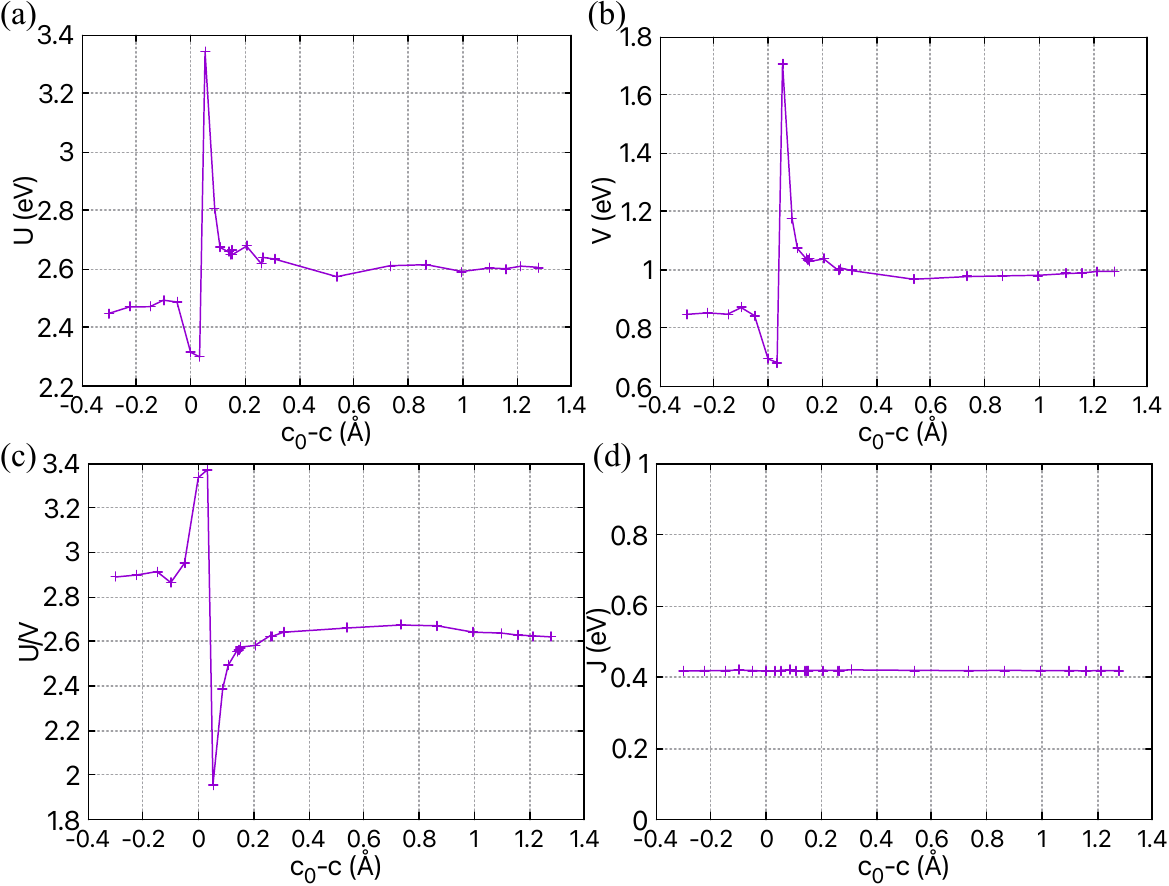}
\caption{The evolution of Coulomb parameters as a function of interlayer spacing in \ch{CsV3Sb5}. A singularity emerges near $c-c_0\approx 0.1$ Å, separating two distinct phases characterized by different values of U and V. Meanwhile, Hund's coupling J remains $\approx 0.42$ eV.}
\label{fig:ccc}
\end{figure*}

FIG.~\ref{fig:ccc} presents the calculated on-site ($U$), nearest-neighbor ($V$), and Hund's coupling ($J$) parameters as functions of interlayer compression ($\Delta c \equiv c-c_0$), where $c$ denotes the c-lattice parameter and $c_0$ represents its value under ambient pressure. Strikingly, $U$ and $V$ exhibit discontinuous transitions resembling metastable step functions, contrasting with the rigid stability of $J$ across the entire range. Specifically, $U$ maintains a baseline value of $2.474$ eV for $\Delta c < 0$ Å but abruptly increases to $2.604$ eV when $\Delta c$ exceeds $0.2$ Å. Remarkably, $V$ undergoes a concurrent enhancement of $\Delta V \approx 0.13$ eV, mirroring $U$'s transition magnitude. These coordinated discontinuities identify a singular critical regime within $0 <\Delta c < 0.2$ Å, rationalizing the anomalous fluctuations observed at ambient pressure. Importantly, despite the concurrent increase in both $U$ and $V$, their ratio $U/V$ demonstrates evident reduction under interlayer compression.

\section{Disscussion and conclusions}

To elucidate this singularity, our findings must be contextualized within the established framework of CDW transitions in \ch{CsV3Sb5}. As demonstrated in Ref.~\cite{denner_analysis_2021}, the KHM exhibits two interwinded ground states: CDO phase ($l = 0$) and chiral CBO phase ($l = 1$). The CDO-CBO crossover is governed by the Coulomb ratio $U/V$, with the CBO phase dominating at reduced $U/V$ values. Crucially, the CBO phase in a 2D-KHM can be lifted to a 3D $2\times2\times 2$ CDW order \cite{ishioka_chiral_2010}, as experimentally resolved via hard-x-ray scattering \cite{li_observation_2021} and scanning tunneling microscopy \cite{xu_three-state_2022}. 

Complementary X-ray scattering study also reveal a coexistence of $2\times2\times 2$ and $2\times2\times 1$ CDW order in \ch{CsV3Sb5} \cite{li_discovery_2022}. Below $p<0.7$ GPa, these orders exhibit degeneracy. Above $p>0.7$ GPa, the degeneracy is lifted because the $2\times2\times 1$ order is suppressed faster than the $2\times2\times 2$ order. Considering the competing nature between CDW and superconductivity in \ch{CsV3Sb5}, it is highly likely the $2\times2\times 1$ CDW order is responsible for the first superconducting dome and $2\times2\times 2$ CDW order is for the second \cite{li_discovery_2022}. 

\begin{figure*}[ht!]
\centering
\includegraphics[width=0.7\textwidth]{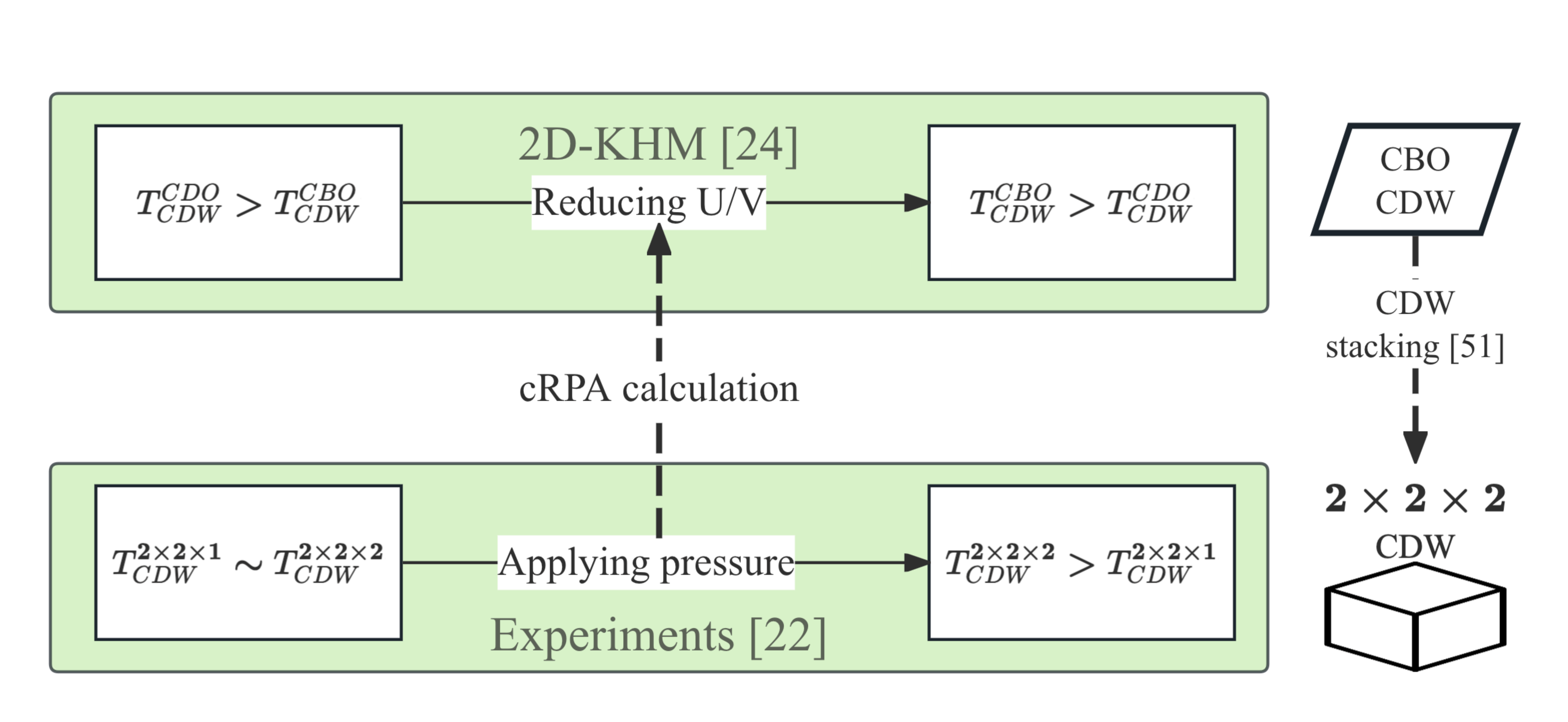}
\caption{Schematic of the relationship between experimental and theoretical study of the CDW phase transition.}
\label{fig:flow}
\end{figure*}

We now schematically compare the 2D-KHM results with experimental observations in FIG.~\ref{fig:flow}. On the one hand, the 2D-KHM study \cite{denner_analysis_2021} indicates that the chiral CBO phase is increasingly favored as the ratio $U/V$ decreases. On the other hand, experimental observations reveal that the chiral $2\times2\times 2$ CDW phase becomes more favored under applied pressure \cite{li_discovery_2022}. Furthermore, the 2D-CBO order can extend into the 3D-$2\times2\times 2$ CDW order through CDW stacking \cite{ishioka_chiral_2010,denner_analysis_2021}. Our cRPA calculations bridge these findings by demonstrating that pressure application can induce an abrupt $U/V$ reduction ($\Delta(U/V) \approx 25 \%$ at $0.2$ GPa). In brief, this reduction could be understood as the signal of a phase transition $X \xrightarrow{\,p\,} 2\times2\times 2$ CDW, where the precursor phase $X$ likely corresponds to a metastable conjoined CDW phase.

\begin{figure*}[hbt!]
\centering
\includegraphics[width=0.8\textwidth]{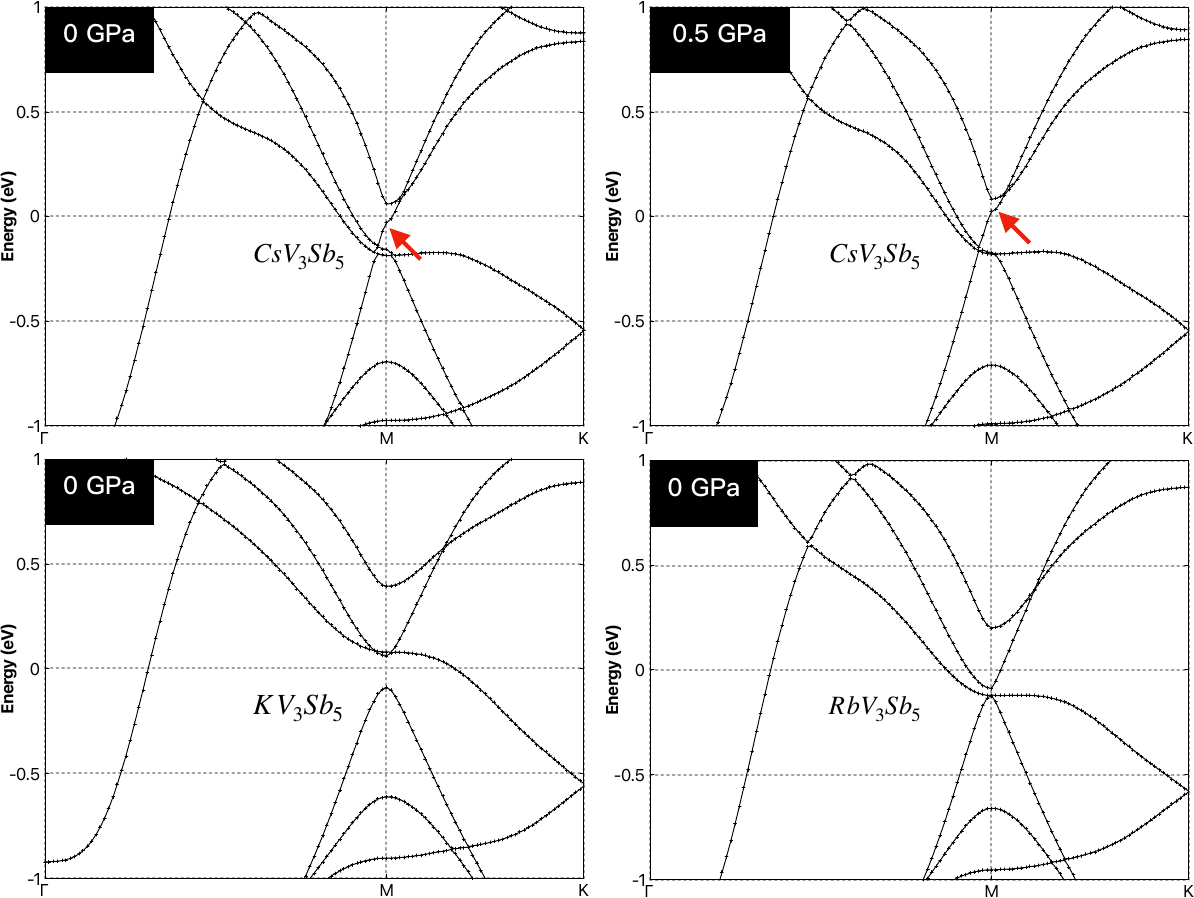}
\caption{GW band structures of \ch{AV3Sb5}. The red arrow highlights the Sb-$5p$ and V-$3d$ mixed VHS in \ch{CsV3Sb5}.}
\label{fig:gwbands}
\end{figure*}

Our systematic analysis reveals a stark dichotomy in pressure response among \ch{AV3Sb5} compounds: while \ch{KV3Sb5} and \ch{RbV3Sb5} maintain pressure-invariant Coulomb interactions ($U$,$V$), \ch{CsV3Sb5} exhibits a pronounced electronic instability at $p \approx 0.2$ GPa characterized by discontinuous jumps in both on-site ($U$) and intersite ($V$) interaction parameters ($\Delta U \approx 0.36$ eV, $\Delta V \approx 0.38$ eV). This critical behavior, absent in isostructural counterparts, strongly suggests a pressure-induced quantum phase transition unique to the Cs variant, likely driven by its interlayer coupling effects because it has the largest interlayer distance. Furthermore, the Hund's $J$ is very stable against pressure in all three compounds, likely plays no role in the pressure-induced quantum phase transition.

To gain deeper insight into the abrupt changes in the Coulomb interactions ($U$, $V$), we revisited the projected band structure of \ch{AV3Sb5} in Fig.~\ref{fig:vhs}. We identify a van Hove singularity (VHS) in \ch{AV3Sb5} slightly above the Fermi level ($E_F$). This VHS, arising from hybridization of Sb-$p$ and V-$d$ orbitals, is sensitive to pressure: its energy shifts upwards with increasing pressure. Notably, it is also potentially linked to the CDW order \cite{PhysRevB.104.144506,li_discovery_2022}. This connection is not coincidental.

Since cRPA calculations inherently include electron self-energy corrections while standard DFT does not, we computed the quasiparticle band structure within the GW approximation \cite{PhysRevB.34.5390,PhysRevB.74.035101,PhysRevB.75.235102} (Fig.~\ref{fig:gwbands}) to ensure consistency with the cRPA methodology. Notably, the GW results position this VHS slightly below $E_F$ at ambient pressure. Furthermore, it crosses $E_F$ under applied pressure. This crossing potentially triggers a Lifshitz transition due to associated changes in Fermi surface topology \cite{2017LTP....43...47V}. In contrast, for \ch{KV3Sb5} and \ch{RbV3Sb5}, the $d$-$p$ hybrid VHS splits energetically at the M point; this splitting, possibly driven by electron-electron interactions, prevents the formation of this specific VHS.

This provides a partial explanation for the peculiar Coulomb interaction behavior in \ch{CsV3Sb5}. Nevertheless, Lifshitz transitions and Coulomb interactions are fundamentally interconnected phenomena. Understanding this interplay is crucial for comprehending electron behavior in Kagome metals. Further investigation focusing on the $d$-$p$ hybrid VHS could provide new insights into this issue.

Building upon the established Coulomb-interaction (U,V) dependence of charge density wave (CDW) orders in kagome metals, our findings demonstrate tentative agreement with both theoretical predictions and experimental observations of intertwined CDW phases in \ch{AV3Sb5} compounds (A = K, Rb, Cs). The calculated pressure-induced variations in $U$ and $V$ parameters provide a new perspective for the unique emergence of double superconducting domes exclusively in \ch{CsV3Sb5}, while remaining absent in its \ch{KV3Sb5} counterpart. Nevertheless, the fundamental role of Coulomb interactions in  governing charge ordering phenomena within quasi-two-dimensional kagome systems remains incompletely understood, particularly regarding: (I) The precise U/V ratio thresholds for CDW phase transitions; (II) The possible dimensional crossover effects on interaction renormalization caused by pressur. These unresolved questions highlight critical gaps in our understanding of correlated electron physics in quasi-2D kagome metals, warranting further investigation.

\appendix

\renewcommand{\thefigure}{A\arabic{figure}}
\setcounter{figure}{0}

\section{K-point convergence of cRPA calculation}

\begin{figure}[hbt!]
\centering
\includegraphics[width=0.4\textwidth]{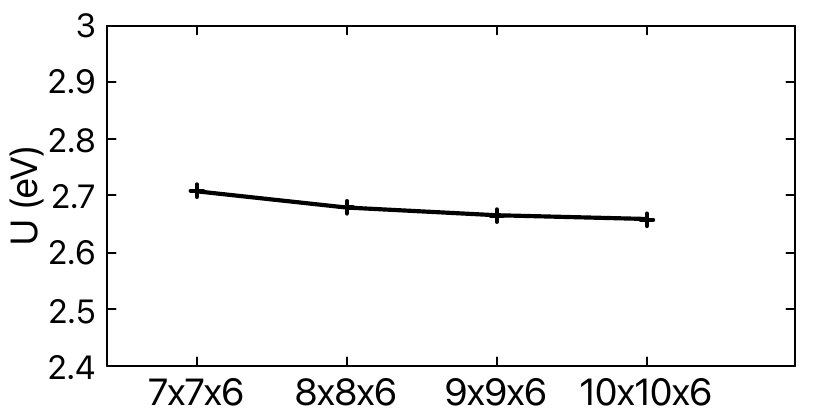}
\caption{The Hubbard U of \ch{CsV3Sb5} at 0.5 GPa with respect to different k-mesh density.}
\label{fig:convergency}
\end{figure}

In cRPA calculations, achieving k-point convergence is crucial for accurate results, as it directly impacts the calculated interaction parameters. Balancing accuracy and computational cost is essential. The goal is to find a mesh that provides sufficient accuracy for our needs without excessive computational time. We calculate the Hubbard $U$ of \ch{CsV3Sb5} at 0.5 GPa with increasing k-mesh density. As shown in FIG.~\ref{fig:convergency}, the k-point convergence can be achieved with $9\times9\times6$ k-mesh.

\vspace{-2em}

\section{Pressure dependence of the bare Coulomb interactions}

\begin{figure}[ht!]
\centering
\includegraphics[width=0.48\textwidth]{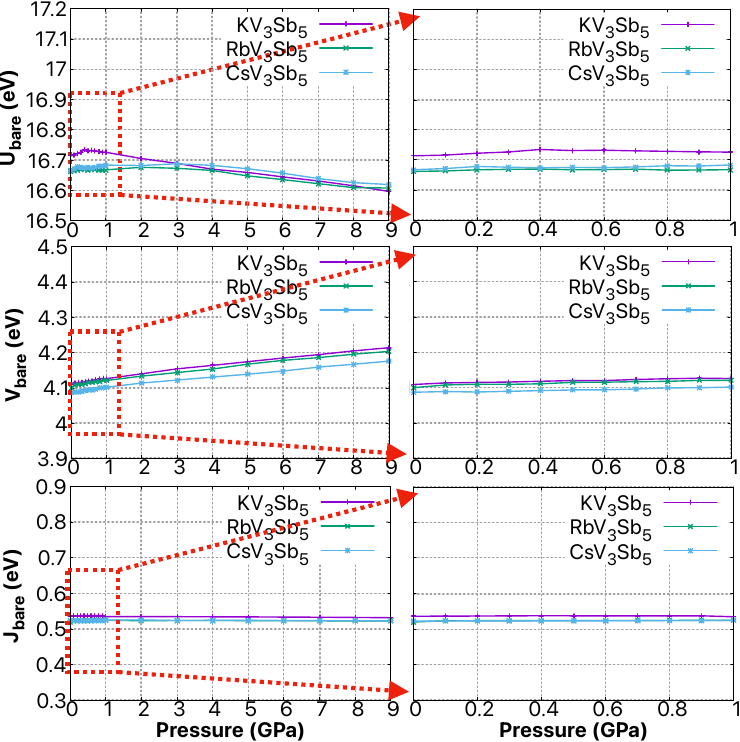}
\caption{The evolution of bare Coulomb interactions as function of pressure. The right column is the enlarged version of the low pressure region.}
\label{fig:bare}
\end{figure}

FIG.~\ref{fig:bare} shows that the bare Coulomb interactions $U_{bare}$, $V_{bare}$ and $J_{bare}$ exhibit no abrupt changes under pressure. This confirms that the sudden changes in the effective $U$ and $V$ are due to the screening effect.

\vspace{-2em}

\begin{acknowledgments}
This work was supported by the Natural Science Foundation of Shaanxi Province (no. 2024JC-YBMS-058).
\end{acknowledgments}

\section*{DATA AVAILABILITY}
Data available from the authors upon reasonable request.

\clearpage
\bibliography{kagome}

\end{document}